\documentclass[12pt,aps,floats]{revtex4}
\usepackage{graphicx}
\usepackage{epsfig}
\usepackage{amssymb}
\usepackage{epstopdf}
\usepackage{amsmath}
\usepackage{graphics}

\newskip\humongous \humongous=0pt plus 1000pt minus 1000pt

\relax

\begin{document}

\title{Thermodynamic phase transition of a Schwarzschild black hole \\with global monopole under GUP}
\bigskip

\author{Yang Chen}
\affiliation{College of Physics Science and Technology,
Shenyang Normal University, Shenyang 110034, China}

\author{Hui-Ling Li}
\email{LHL51759@126.com} \affiliation{College of Physics Science and Technology,
Shenyang Normal University, Shenyang 110034, China}

\begin{abstract}
\textbf{Abstract:}

In this paper, considering the influence of the principle of generalized uncertainty (GUP), the phase transition of a Schwarzschild black hole with global monopole is discussed. First, we use the generalized Dirac equation to obtain corrected Hawking temperature, local temperature, black hole residue, black hole entropy, thermal capacity, and other thermodynamic quantities. Then, we use images to analyze the effects of generalized uncertainty parameters on phase transitions and the effects of magnetic monopole parameters on phase transitions. Finally, we study the thermodynamic stability and phase change structure under the influence of the principle of generalized uncertainty. The results show that, in the black hole with global monopole, there are first-order and second-order phase transitions. In addition, the general uncertainty parameters and monopole parameters will affect the black hole residues.


\end{abstract}

\maketitle

\section{Introduction}

In general relativity, black holes are the most surprising objects. In classical (non-quantum) physics, any matter (including light) entering a black hole cannot escape. With the continuous exploration of black holes, scientists have become extremely interested in the thermodynamics of black holes. In 1972, Beckenstein proposed the ``hairless theorem'' for black holes. He believed that after the star collapsed into a black hole, only the three basic conserved quantities of mass, angular momentum and electric charge continued to play a role. Some other factors (``hair'') disappeared after entering the black hole. This theorem was later rigorously proved by Hawking and others \cite{1,2}. Bekenstein also conceived that the second law of thermodynamics should be universally valid \cite{3,4}, and he believed that black holes should have the entropy proportional to their surface area.

Black holes have rich critical phenomena and phase structures. In 1983, Hawking et al. \cite{5} found through calculation that in the Schwarzschild--anti de Sitter space, there was a phase transition between the stable large black hole phase and the hot gas phase, which is called Hawking--Page phase transition. Through this discovery, physicists have a thorough understanding of the relationship between black hole physics and traditional thermodynamic systems, and opened the door to the study of black hole phase transitions. The location of the Hawking--Page transition between the charged black hole and the pure anti-de Sitter space--time was clearly discuss in 1999 \cite{6}. It is studied whether the Hawking--Page transition can occur in three dimensions in 2005 \cite{7}. The thermodynamic phase transitions of high--dimensional single--spin asymptotic AdS black holes in a canonical (fixed J) set of extended phase space are studied in 2013 \cite{8}. In addition, the criticality index of charged AdS black hole is also calculated, and the results show that they are consistent with the criticality index of the van der Waals system \cite{9}.

In the vacuum phase transition of the early universe, the breakdown of local or global scale symmetry produced topological defects such as domain walls, cosmic strings, and monopoles \cite{10,11}. It was proved in \cite{12} that during the phase transition of the universe, the global monopole charge spontaneously breaks into U (1) with the global O (3) symmetry. In 1994, H. W. Yu et al. studied the thermodynamics of a static spherically symmetric black hole--global unipolar system through two methods, surface gravity and Euclidean path integration \cite{13}. In \cite{14}, Dadhich et al. studied the effect of global monopolar charge on particle orbits and Hawking radiation, and corrected the global monopole black hole Hawking radiation in \cite{15}. In addition, Kimet and Gordana proposed the quantum tunneling effect of the Dirac magnetic monopole in a black hole with an integral magnetic monopole \cite{16}. In 2016, researchers studied the mass fraction of the unipolar parameter and found that it can suppress the maximum accretion rate \cite{17}.

On the other side, the generalized uncertainty principle has been applied to different studies. The form of GUP is
\begin{eqnarray} \label{1}
\Delta x\Delta p \geqslant \frac{\hbar }{2}[1 + \beta {(\Delta p)^2}],
\end{eqnarray}
where $ \Delta x $ and $ \Delta p $ represent the uncertainty of position and momentum, respectively. GUP parameter $\beta  = {\beta _0}l_{_p}^2/{\hbar ^2} = {\beta _0}/{M_p}{c^2}$, $ l_{p}$ is the Planck length ($ \approx {10^{ - 35}}$), $ M_{p} $ is Planck mass and $ \beta_{0} $ is a dimensionless parameter. GUP is a low--energy approximation theory. There are three general types of GUP. One of them is GUP with minimum position \cite{18} and the other two are GUP with maximum momentum and minimum position \cite{19,20}, and GUP with minimum position, minimum momentum, and maximum momentum \cite{21}. Adler et al. \cite{22} found that after considering the corrected of GUP, the black holes will not evaporate completely, leaving residual mass. In black hole physics, many meaningful conclusions are drawn by applying different forms of GUP \cite{23,24,25}. Based on the GUP, we will explore the phase transition of a Schwarzschild black hole with a global monopole.

The structure of this paper is as follows: In Section 2, we review the modified Dirac equation and use the Dirac equation to derive the relationship between local temperature and mass in a Schwarzschild black hole with the global monopole. The effects of GUP on phase transitions are discussed in Section 3. We finally concluded.

\section{Quantum tunneling and local temperature based on GUP}
In this section, firstly, we consider the effects of quantum gravity and discuss the tunneling behavior of fermions through the Schwarzschild black hole with the global monopole event horizon. In curved space-time, the Generalized Dirac equation can be expressed as \cite{26}
\begin{eqnarray} \label{2}
 [i{\gamma ^0}{\partial _0} + i{\gamma ^i}{\partial _i}(1 - \beta {m^2}) + i{\gamma ^i}\beta {\hbar ^2}({\partial _j}{\partial ^j}){\partial _i} + \frac{m}{\hbar }(1 + \beta {\hbar ^2}{\partial _j}{\partial ^j} - \beta {m^2})\nonumber  \\
 + i{\gamma ^\mu }{\Omega _\mu }(1 + \beta {\hbar ^2}{\partial _j}{\partial ^j} - \beta {m^2})]\psi  = 0.
\end{eqnarray}
The curved space-time metric of Schwarzschild black hole with the global monopole is
\begin{eqnarray} \label{3}
d{S^2} =  - {\rm{f}}\left( r \right)d{t^2} + \frac{1}{{g\left( r \right)}}d{r^2} + a{r^2}\left( {d{\theta ^2} + {{\sin }^2}\theta d{\phi ^2}} \right),
\end{eqnarray}
where $a = 1 - 8\pi {\eta ^2}$ and
\begin{eqnarray} \label{4}
f\left( r \right) = g\left( r \right) = 1 - \frac{{2GM}}{r}.
\end{eqnarray}
The mass of black hole is expressed by $ M $. We have set $G = c = 1$ . This metric is not asymptotic flat space time due to the existence of global magnetic monopoles. When $ f\left( {{r_ + }} \right){\rm{ = 0}}$, event horizon ${r_ + } = 2GM$ can be obtained. The fermion¡¯s movement follows the generalized Dirac equation (2). Therefore, in order to describe the motion, we assume that the wave function state is
\begin{eqnarray} \label{5}
\psi  = \left( \begin{array}{l}
 A \\
 0 \\
 B \\
 0 \\
 \end{array} \right)\exp \left( {\frac{i}{\hbar }I\left( {t,r,\theta ,\phi } \right)} \right),
\end{eqnarray}
where $I$ is the action of the emitted fermions, $A$ and $B$ are functions of coordinates, that is $A = A\left( {t,r,\theta ,\phi } \right),B = B\left( {t,r,\theta ,\phi } \right)$.
Then the ${\gamma ^\mu }$ matrices are given by
\begin{eqnarray} \label{6}
 {\gamma ^t} = \frac{1}{{\sqrt {f\left( r \right)} }}\left( {\begin{array}{*{20}{c}}
   i & 0  \\
   0 & { - i}  \\
\end{array}} \right),{\gamma ^\theta } = \sqrt {{g^{\theta \theta }}} \left( {\begin{array}{*{20}{c}}
   0 & {{\sigma ^1}}  \\
   {{\sigma ^1}} & 0  \\
\end{array}} \right), \nonumber  \\
 {\gamma ^r} = \sqrt {g\left( r \right)} \left( {\begin{array}{*{20}{c}}
   0 & {{\sigma ^3}}  \\
   {{\sigma ^3}} & 0  \\
\end{array}} \right),{\gamma ^\phi } = \sqrt {{g^{\phi \phi }}} \left( {\begin{array}{*{20}{c}}
   0 & {{\sigma ^2}}  \\
   {{\sigma ^2}} & 0  \\
\end{array}} \right),
\end{eqnarray}
where $\sqrt {{g^{\theta \theta }}}  = \frac{1}{{\sqrt a r}},\sqrt {{g^{\phi \phi }}}  = \frac{1}{{\sqrt a r\sin \theta }}$. ${\sigma ^{\rm{i}}}$¡¯s are the Pauli matrices with $i = 1,2,3$.
Substituting the wave function and gamma matrices into the generalized Dirac equation (2), ignoring the contributions from $\partial A$, $\partial B$ and ignoring high-order terms of $\hbar $ , we finally receive the following equations of motion
\begin{eqnarray} \label{7}
&& - iA\frac{1}{{\sqrt f }}{\partial _t}I - B\left( {1 - \beta {m^2}} \right)\sqrt g {\partial _r}I - Am\beta [{g^{rr}}{({\partial _r}I)^2} + {g^{\theta \theta }}{({\partial _\theta }I)^2} + {g^{\phi \phi }}{({\partial _\phi }I)^2}] \nonumber  \\
&&+ B\beta \sqrt g {\partial _r}I[{g^{rr}}{({\partial _r}I)^2} + {g^{\theta \theta }}{({\partial _\theta }I)^2} + {g^{\phi \phi }}{({\partial _\phi }I)^2}] + Am\left( {1 - \beta {m^2}} \right) = 0,
\end{eqnarray}
\begin{eqnarray} \label{8}
&&iB\frac{1}{{\sqrt f }}{\partial _t}I - A\left( {1 - \beta {m^2}} \right)\sqrt g {\partial _r}I - Bm\beta [{g^{rr}}{({\partial _r}I)^2} + {g^{\theta \theta }}{({\partial _\theta }I)^2} + {g^{\phi \phi }}{({\partial _\phi }I)^2}]  \nonumber  \\
&&+ A\beta \sqrt g {\partial _r}I[{g^{rr}}{({\partial _r}I)^2} + {g^{\theta \theta }}{({\partial _\theta }I)^2} + {g^{\phi \phi }}{({\partial _\phi }I)^2}] + Bm\left( {1 - \beta {m^2}} \right) = 0,
\end{eqnarray}
\begin{eqnarray} \label{9}
  &&A\left( {1 - \beta {m^2}} \right)\sqrt {{g^{\theta \theta }}} {\partial _\theta }I - A\beta \sqrt {{g^{\theta \theta }}} {\partial _\theta }I[{g^{rr}}{({\partial _r}I)^2} + {g^{\theta \theta }}{({\partial _\theta }I)^2} + {g^{\phi \phi }}{({\partial _\phi }I)^2}] \nonumber  \\
  &&+ iA\left( {1 - \beta {m^2}} \right)\sqrt {{g^{\phi \phi }}} {\partial _\phi }I - iA\beta \sqrt {{g^{\phi \phi }}} {\partial _\phi }I[{g^{rr}}{({\partial _r}I)^2} + {g^{\theta \theta }}{({\partial _\theta }I)^2} + {g^{\phi \phi }}{({\partial _\phi }I)^2}] = 0,
\end{eqnarray}
\begin{eqnarray} \label{10}
&&B\left( {1 - \beta {m^2}} \right)\sqrt {{g^{\theta \theta }}} {\partial _\theta }I - B\beta \sqrt {{g^{\theta \theta }}} {\partial _\theta }I[{g^{rr}}{({\partial _r}I)^2} + {g^{\theta \theta }}{({\partial _\theta }I)^2} + {g^{\phi \phi }}{({\partial _\phi }I)^2}] \nonumber  \\
&&+ iB\left( {1 - \beta {m^2}} \right)\sqrt {{g^{\phi \phi }}} {\partial _\phi }I - iB\beta \sqrt {{g^{\phi \phi }}} {\partial _\phi }I[{g^{rr}}{({\partial _r}I)^2} + {g^{\theta \theta }}{({\partial _\theta }I)^2} + {g^{\phi \phi }}{({\partial _\phi }I)^2}] = 0.
\end{eqnarray}
It is difficult to solve the action from the above equations directly. Following the standard process, we carry out the separation of variables as follows
\begin{eqnarray} \label{11}
I =  - ({\rm{1 - 8}}\pi {\eta ^{\rm{2}}})\omega {\rm{t + W}}\left( r \right) + \Theta \left( {\theta ,\phi } \right),
\end{eqnarray}
where $\omega $ is the energy of emitted fermions. We substitute eqn. (11) into eqn.s (7)-(10) and observe the last two equations. After dividing by $A$ and $B$ respectively, they are found to be same, so we can find
\begin{eqnarray} \label{12}
\left( {\sqrt {{g^{\theta \theta }}} {\partial _\theta }\Theta  + i\sqrt {{g^{\phi \phi }}} {\partial _\phi }\Theta } \right)\left[ {\beta {g^{rr}}{{\left( {{\partial _r}W} \right)}^2} + \beta {g^{\theta \theta }}{{\left( {{\partial _\theta }\Theta } \right)}^2} + \beta {g^{\phi \phi }}{{\left( {{\partial _\phi }\Theta } \right)}^2} - \left( {1 - \beta {m^2}} \right)} \right] = 0.
\end{eqnarray}
In the above equation, $\beta $ is a small quantity and represents the effects from quantum gravity. The value in the square bracket can¡¯t vanish. Therefore eqn. (12) is simplified as follow
\begin{eqnarray} \label{13}
\sqrt {{g^{\theta \theta }}} {\partial _\theta }\Theta  + i\sqrt {{g^{\phi \phi }}} {\partial _\phi }\Theta=0.
\end{eqnarray}
In the previous study, $\Theta $ has a complex function. However, it has no effect on the tunneling rate. Then by canceling $A$ and $B$, eqn. (7) and eqn. (8) are identical and written as
\begin{eqnarray} \label{14}
{B_6}{\left( {{\partial _r}W} \right)^6} + {B_4}{\left( {{\partial _r}W} \right)^4} + {B_2}{\left( {{\partial _r}W} \right)^2} + {B_0} = 0,
\end{eqnarray}
where
\begin{eqnarray} \label{15}
 &&{B_6} = {\beta ^2}{g^3}f ;\nonumber  \\
 &&{B_4} = \beta {g^2}f({m^2}\beta  + 2\beta Q - 2) \nonumber ; \\
 &&{B_2} = gf[{(1 - \beta {m^2})^2} + \beta (2{m^2} - 2{m^4}\beta  - 2Q + \beta {Q^2})] ;\\
 &&{B_0} =  - {m^2}{(1 - \beta {m^2} - \beta Q)^2}f - {\omega ^2}{(1 - 8\pi {\eta ^2})^2} \nonumber  ;\\
&&Q = {g^{\theta \theta }}{\left( {{\partial _\theta }\Theta } \right)^2} + {g^{\phi \phi }}{\left( {{\partial _\phi }\Theta } \right)^2} \nonumber.
\end{eqnarray}
Ignoring the higher order terms of $\beta $ and solving the above equations at the event horizon, we get
\begin{eqnarray} \label{16}
 {W_ \pm } &&=  \pm \int {dr\frac{1}{{\sqrt {fg} }}} \sqrt {{\omega ^2} - {m^2}f + 2\beta {m^4}f} \left[ {1 + \beta \left( {{m^2} + \frac{{{\omega ^2}}}{f}} \right)} \right] \nonumber \\
 &&=  \pm i\pi 2M\omega (1 - 8\pi {\eta ^2})(1 + \frac{1}{2}\beta (3{m^2} + 4{(1 - 8\pi {\eta ^2})^2}{\omega ^2})) + \text{real part},
\end{eqnarray}
where ${ +  \mathord{\left/{\vphantom { +   - }} \right.\kern-\nulldelimiterspace}  - }$ signs represent the outgoing/ingoing solutions. The real part is independent of the tunneling rate. Therefore, the tunneling rate is written as
\begin{eqnarray} \label{17}
\Gamma  &&= \exp ( - 4{\mathop{\rm Im}\nolimits} {W_ + })\nonumber \\
 &&= \exp \left\{ { - 8\pi M\omega (1 - 8\pi {\eta ^2})\left\{ {1 + \frac{1}{2}\beta \left[ {3{m^2} + 4{{(1 - 8\pi {\eta ^2})}^2}{\omega ^2}} \right]} \right\}} \right\},
\end{eqnarray}
which shows Hawking temperature is written as
\begin{eqnarray} \label{18}
 T &&= {\left\{ {8M\pi (1 - 8\pi {\eta ^2})\left\{ {1 + \frac{1}{2}\beta \left[ {3{m^2} + 4{{(1 - 8\pi {\eta ^2})}^2}{\omega ^2}} \right]} \right\}} \right\}^{ - 1}} \nonumber \\
 && \simeq {T_H}\left\{ {1 - \frac{1}{2}\beta \left[ {3{m^2} + 4{{(1 - 8\pi {\eta ^2})}^2}{\omega ^2}} \right]} \right\}.
\end{eqnarray}
Then, we further discuss on the local temperature based on GUP.
According to \cite{27,28}, in order to obtain a local thermodynamic quantity, the Schwarzschild black hole with global monopole needs to be located at the center of the spherical cavity of radius $r$. Where ${T_H} = \frac{1}{{4\pi {r_ + }}}$ is Black hole's original Hawking temperature, where ${{\rm{r}}_ + } = 2GM$. It is worth noting that the ADM quality of the magnetic monopole is $(1 - 8\pi {\eta ^2})M$ \cite{29}. So, Hawking temperature can be written as
\begin{eqnarray} \label{19}
{T_H} = \frac{1}{{8\pi (1 - 8\pi {\eta ^2})GM}}.
\end{eqnarray}
Local temperature in the Schwarzschild black hole can be written as follows
\begin{eqnarray} \label{20}
{T_{Local}} = \frac{T}{{\sqrt {f(r)} }} = \frac{1}{{\sqrt {1 - \frac{{2M}}{r}} }}\frac{1}{{8\pi G(1 - 8\pi {\eta ^2})M}}\left\{ {1 - \frac{1}{2}\beta \left[ {3{m^2} + 4{{(1 - 8\pi {\eta ^2})}^2}{\omega ^2}} \right]} \right\}.
\end{eqnarray}
In Hawking radiation, in order to facilitate the calculation, we can get the lower limit of the energy of the released particles from the saturated form of the uncertainty principle,
\begin{eqnarray} \label{21}
\omega ' \ge \frac{\hbar }{{\Delta x}}.
\end{eqnarray}
It is noteworthy that $\omega '$ in the magnetic monopole black hole is $(1 - 8\pi {\eta ^2})\omega $. In the vicinity of the black hole event horizon, the uncertainty can be used in the radius of the black hole \cite{30}. So we can be obtain
\begin{eqnarray} \label{22}
\Delta x \approx {r_{BH}} = {r_ + }.
\end{eqnarray}
From equations (21) and (22), we have
\begin{eqnarray} \label{23}
\omega ' = \frac{\hbar }{{{r_ + }}}.
\end{eqnarray}
So the local temperature can be written as
\begin{eqnarray} \label{24}
{T_{Local}} = \frac{1}{{\sqrt {1 - \frac{{2M}}{r}} }}\frac{1}{{8\pi G(1 - 8\pi {\eta ^2})M}}\left\{ {1 - \frac{1}{2}\beta \left[ {3{m^2} + {{\left( {\frac{\hbar }{M}} \right)}^2}} \right]} \right\}.
\end{eqnarray}
Based on the above equation, we can to find out the critical value of phase transition. When considering the cavity of the radius r as a constant amount, and setting $8\pi {\eta ^2}{\rm{ = }}X$, the following three critical values can be calculated by the following equations:
\begin{eqnarray} \label{25}
{\left( {\frac{{\partial {T_{Local}}}}{{\partial M}}} \right)_{\eta ,{\rm{r}}}} = {\left( {\frac{{{\partial ^2}{T_{Local}}}}{{\partial {M^2}}}} \right)_{\eta ,{\rm{r}}}}=0.
\end{eqnarray}
Setting $r = 10,\hbar  = 1$, and substituting equation (24) into equation (25), critical mass, critical GUP parameters, and critical local temperature can be obtained as follows ${\beta _{C{\rm{r}}}} = 2.52068,{M_{Cr}} = 2.59934,{T_{Cr}} = 0.019964$.
According to Eq. (24), the relationship between local temperature versus mass as a function can be shown as Figures 1-3.

\begin{figure}[h]
\centering
\includegraphics[width=0.6\textwidth]{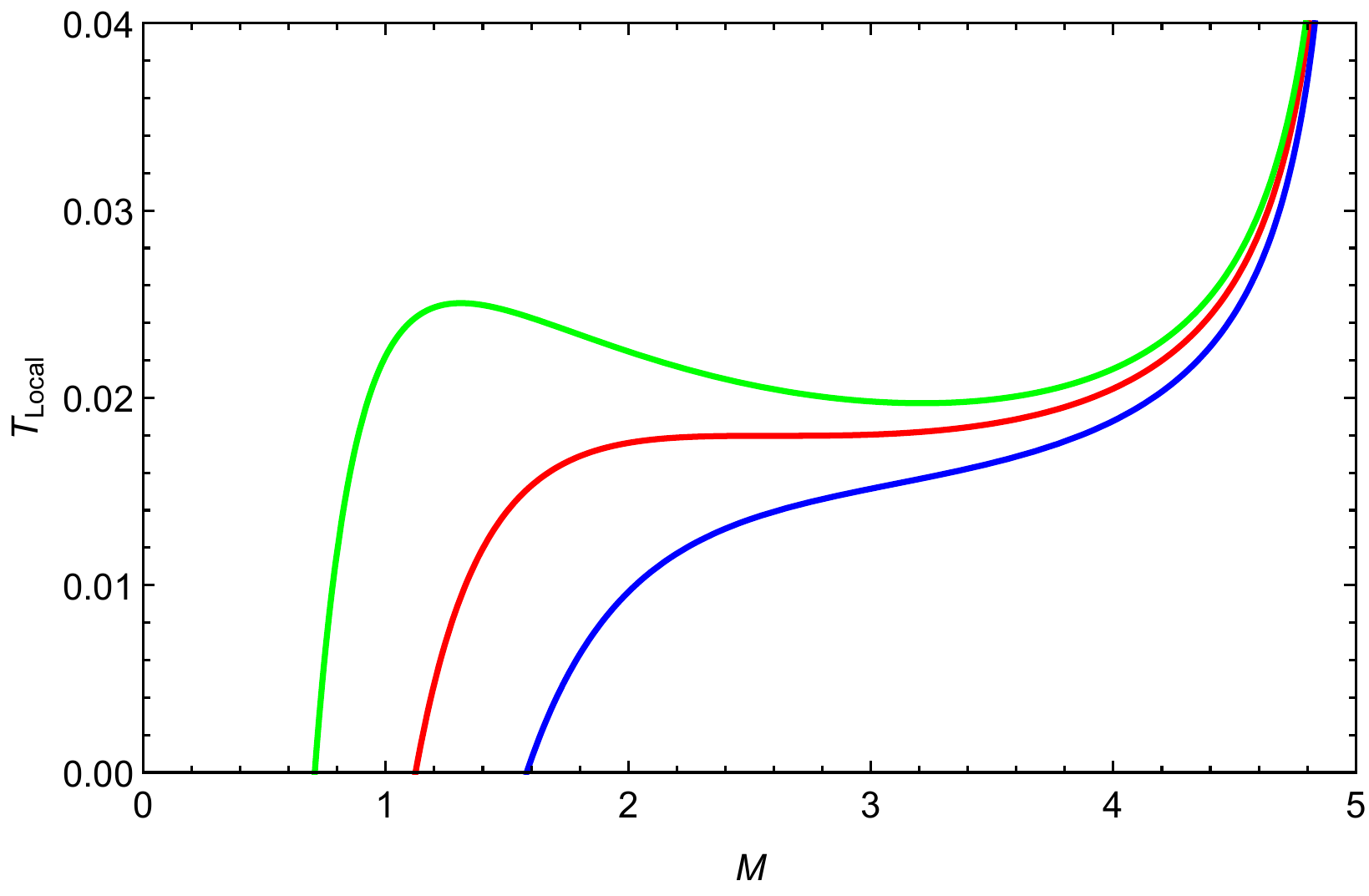}
\caption{\small Fixed $G=1$, $r=10$ and $X = 0.1$, plot a graph of the relationship between ${T_{L{\rm{ocal}}}}$ and $M$ for $\beta  = 1 < {\beta _C}$ (top), $\beta {\rm{ = }}{\beta _C}$(intermediate), and $\beta  = 5 > {\beta _C}$(bottom).}
\end{figure}

Obviously, if $\beta  < {\beta _C}$, the image curve will fluctuate, which indicates that there may be a phase change. As $\beta $ increases, the image curve monotonically increases and the inflection point disappears, which indicates that no phase change may occur.
\begin{figure}[h]
\centering
\includegraphics[width=0.6\textwidth]{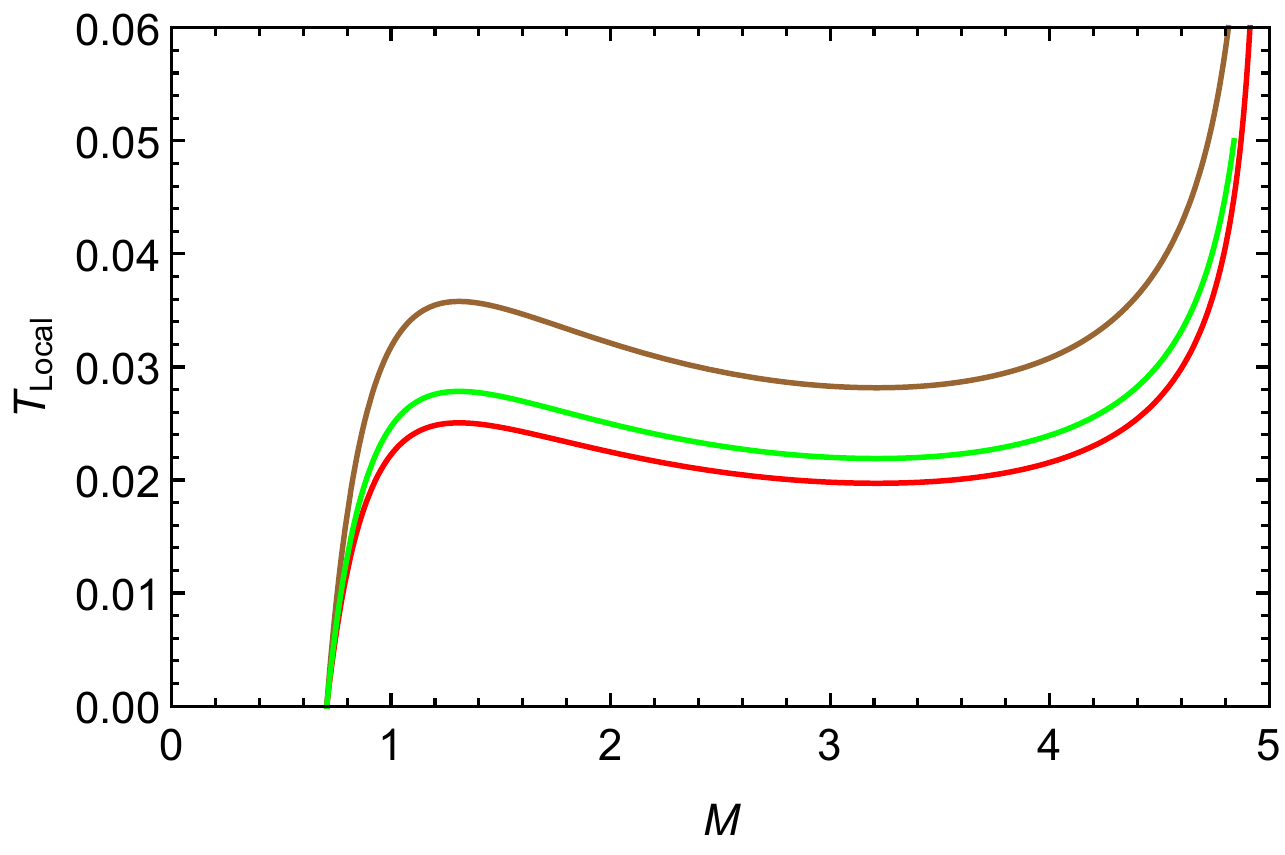}
\caption{\small Fixed $G=1$, $r=10$ and $\beta {\rm{ = }}1$, plot a graph of the relationship between ${T_{L{\rm{ocal}}}}$ and $M$. The curve is the image from top to bottom for $X = {\rm{ }}0.{\rm{3}},0.{\rm{1}},0.05$.}
\end{figure}

The trend of the image is similar to that in Figure 1, so a phase change occurs.

\begin{figure}[h]
\centering
\includegraphics[width=0.6\textwidth]{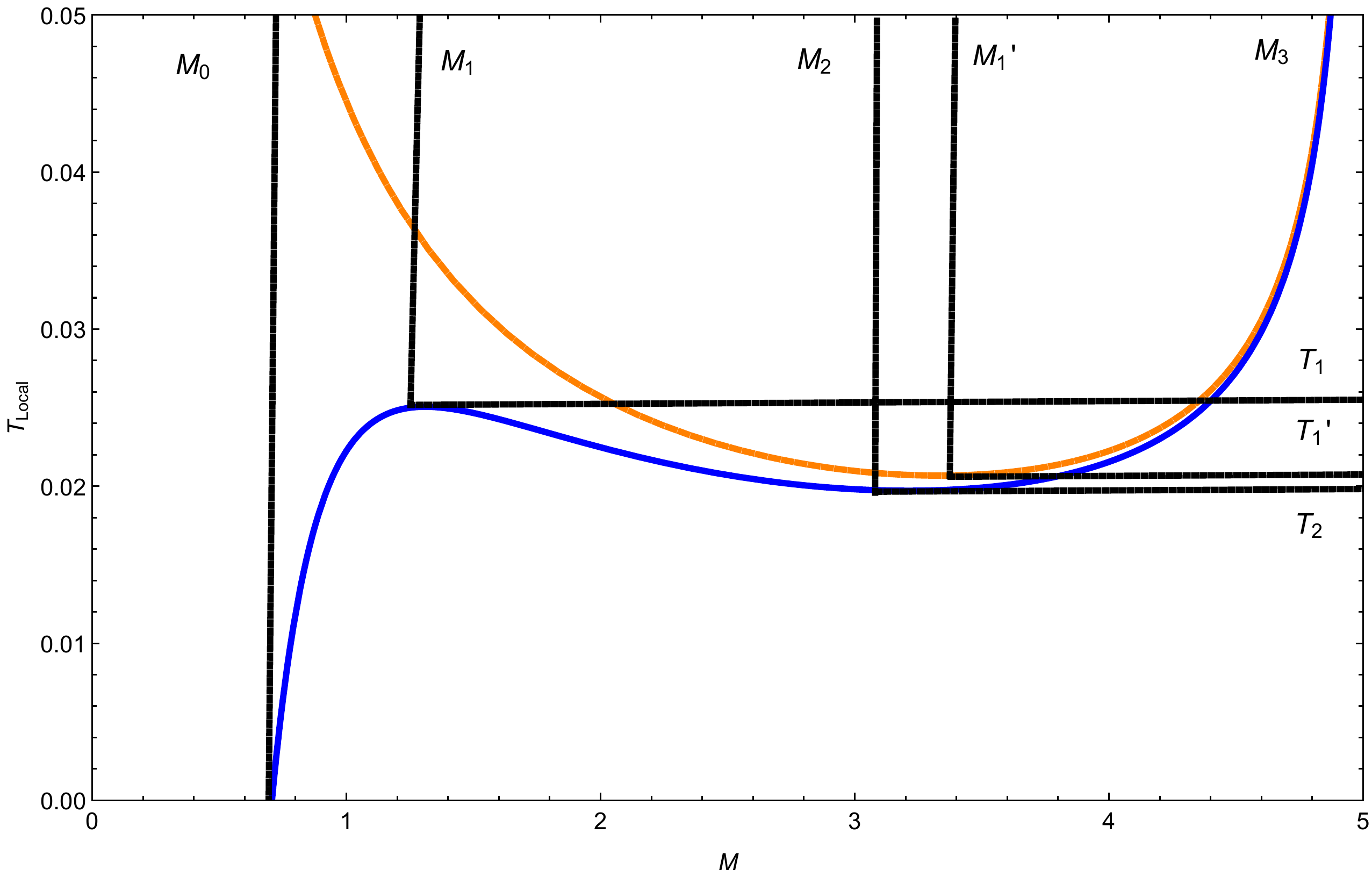}
\caption{\small Fixed $G=1$, $r=10$ and $X = 0.1$, plot a graph of the relationship between ${T_{L{\rm{ocal}}}}$ and $M$. The curve is the image from top to bottom for $\beta {\rm{ = 0}}$ and $\beta {\rm{ = 1}}$.}
\end{figure}

Form Figure 3, the relationship between the original local temperature (for $\beta {\rm{ = }}0$) and mass and the relationship between local temperature (for $\beta {\rm{ = }}1$) and mass can be obtained respectively. When $\beta {\rm{ = }}1$, the evaporation process of the Schwarzschild black hole with global monopole is generally divided into three stages. In the first stage (${{\rm{M}}_{\rm{2}}} \le {\rm{M}} \le {{\rm{M}}_{\rm{3}}}$), the local temperature is reduced by the evaporation process. In the second stage (${{\rm{M}}_{\rm{1}}} \le {\rm{M}} \le {{\rm{M}}_{\rm{2}}}$), the local temperature decreases with the mass increases. In the final stage (${{\rm{M}}_0} \le {\rm{M}} \le {{\rm{M}}_{\rm{1}}}$), because the negative temperature violates the laws of thermodynamics, the mass of the black hole cannot be less than ${{\rm{M}}_0}$. When the mass is close to ${{\rm{M}}_0}$, the local temperature becomes 0, which results in a residual black hole, that is ${M_0} = {M_{res}} = \sqrt {\frac{{{\hbar ^2}\beta }}{2}} $.
Therefore, a large black hole (LBH) of ${{\rm{M}}_{\rm{2}}} \le {\rm{M}} \le {{\rm{M}}_{\rm{3}}}$, an intermediate black hole (IBH) of ${{\rm{M}}_{\rm{1}}} \le {\rm{M}} \le {{\rm{M}}_{\rm{2}}}$, and a small black hole (SBH) of ${{\rm{M}}_0} \le {\rm{M}} \le {{\rm{M}}_{\rm{1}}}$ can be defined.

\section{The effect of GUP on phase transition}
According to the first law of thermodynamics, the local energy of the Schwarzschild black hole with global monopole can be given by
\begin{eqnarray} \label{26}
{\left( {\frac{{\partial {T_{Local}}}}{{\partial M}}} \right)_{\eta ,{\rm{r}}}} = {\left( {\frac{{{\partial ^2}{T_{Local}}}}{{\partial {M^2}}}} \right)_{\eta ,{\rm{r}}}}=0.
\end{eqnarray}
That is
\begin{eqnarray} \label{27}
 {E_{Local}} &&= \mathop \smallint \limits_{(1 - 8\pi {\eta ^2}){M_0}}^{(1 - 8\pi {\eta ^2})M} \frac{{{T_{Local}}}}{{{T_H}}}d(1 - 8\pi {\eta ^2})M \nonumber \\
  &&= \frac{{(1 - 8\pi {\eta ^2})}}{{2( - 1 + 8\pi {\eta ^2})}}\sqrt {\frac{{2( - 1 + 8\pi {\eta ^2})M + r}}{r}} [r(2 - 3{m^2}\beta ) + \frac{{\beta {\hbar ^2}}}{{( - 1 + d)M}}] \nonumber \\
  &&- \frac{{(1 - 8\pi {\eta ^2})}}{{2( - 1 + 8\pi {\eta ^2})}}\left[ {r(2 - 3{m^2}\beta ) + \frac{{\sqrt 2 \sqrt {\beta {\hbar ^2}} }}{{ - 1 + 8\pi {\eta ^2}}}} \right]\sqrt {\frac{{r + \sqrt 2 ( - 1 + 8\pi {\eta ^2})\sqrt {\beta {\hbar ^2}} }}{r}}  \nonumber \\
  &&- \frac{{(1 - 8\pi {\eta ^2})}}{{2( - 1 + 8\pi {\eta ^2})}}\frac{{2\beta {\hbar ^2}ArcTanh[\sqrt {\frac{{2( - 1 + 8\pi {\eta ^2})M + r}}{r}} ]}}{r} \nonumber \\
  &&+ \frac{{(1 - 8\pi {\eta ^2})}}{{2( - 1 + 8\pi {\eta ^2})}}\frac{{2\beta {\hbar ^2}ArcTanh[\sqrt {\frac{{r + \sqrt 2 ( - 1 + 8\pi {\eta ^2})\sqrt {\beta {\hbar ^2}} }}{r}} ]}}{r} .
\end{eqnarray}
Meanwhile, we can obtain heat capacity
\begin{eqnarray} \label{28}
 C &&= \frac{{\partial {E_{Local}}}}{{\partial {T_{Local}}}} = \frac{{\frac{{\partial {E_{Local}}}}{{\partial M}}}}{{\frac{{\partial {T_{Local}}}}{{\partial M}}}} \nonumber \\
  &&= \frac{{8G{M^2}\pi \sqrt {1 - \frac{{2M}}{r}} ( - 2M + r)({{( - 1 + 8\pi {\eta ^2})}^2}{M^2}( - 2 + 3{m^2}\beta ) + \beta {\hbar ^2})}}{{\sqrt {\frac{{2( - 1 + 8\pi {\eta ^2})M + r}}{r}} ({M^2}r(2 - 3{m^2}\beta ) + {M^3}( - 6 + 9{m^2}\beta ) + 7M\beta {\hbar ^2} - 3r\beta {\hbar ^2})}}.
 \end{eqnarray}
The heat capacity associated with the quality of the Schwarzschild black hole with global magnetic monopoles is shown in Figure 4.

\begin{figure}[h]
\centering
\includegraphics[width=0.6\textwidth]{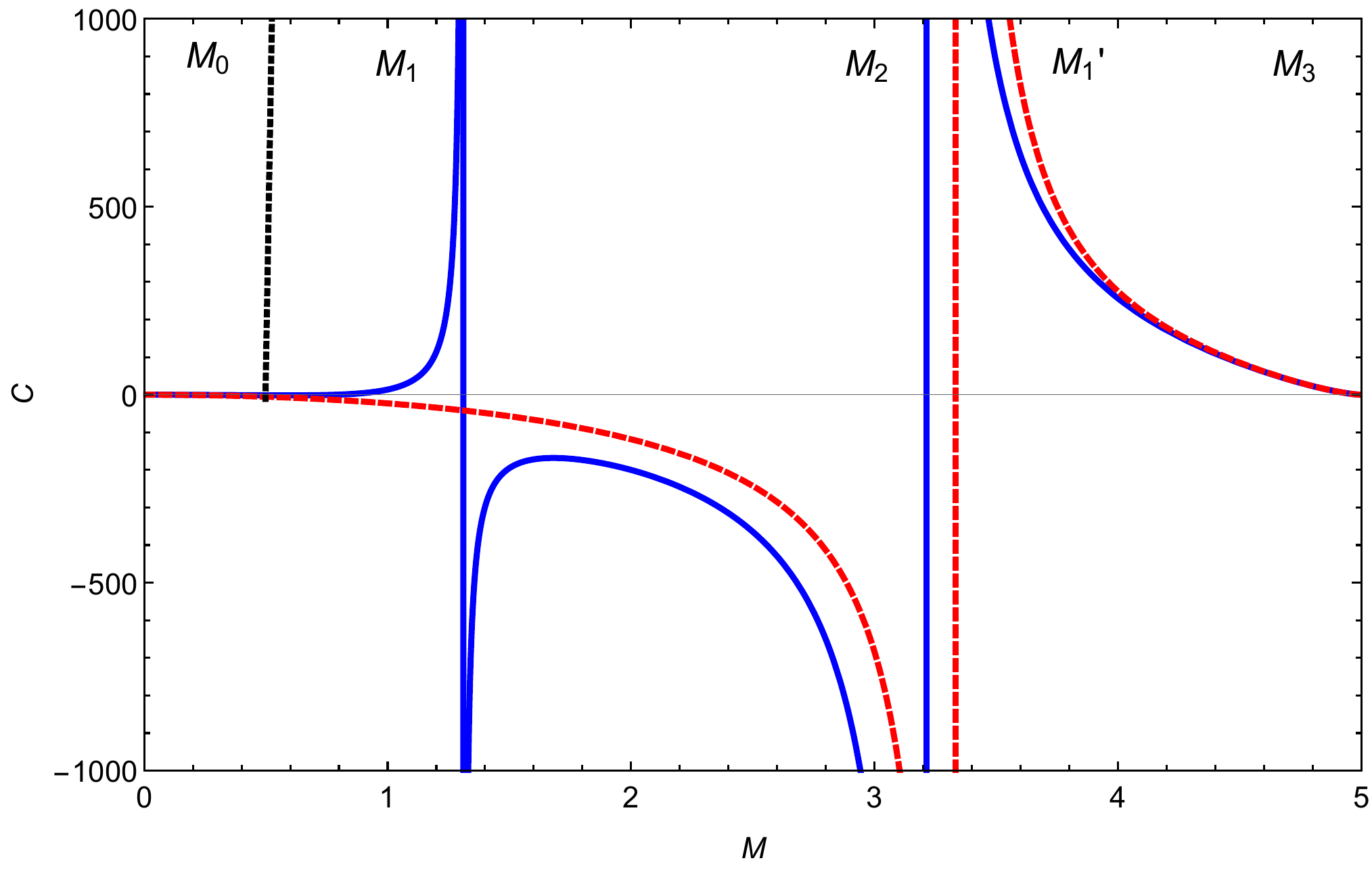}
\caption{\small Fixed $G=1$, $r=10$ and $X = 0.1$, plot the relationship between $C$ and $M$. The figure is the $C-M$ image for $\beta {\rm{ = 0}}$ and $\beta {\rm{ = 1}}$ respectively.}
\end{figure}

As shown in Fig. 4, the solid blue line is an image of the thermal capacity of a black hole and the mass change of the black hole when the GUP parameter is taken into account. It can be seen from the figure that there are two stable regions: ${M_0} \le M \le {M_1}$ and ${M_2} \le M \le {M_3}$.  While when ${M_1} \le M \le {M_2}$, the black hole is in an unstable state. Since the thermal capacity is negative, the red dotted line represents an image of the original heat capacity as a function of the mass of the black hole. When the mass is zero, the thermal capacity also approaches 0. As the mass increases, a divergence point appears at ${M_2}$, the thermal capacity on the left side of the divergence point is negative, the black hole is unstable, the thermal capacity on the right side of the divergence point is positive, and the black hole is relatively stable. It can be clearly seen that the images diverge at ${M_1}$  and ${M_2}$ , proving the existence of a phase transition.
Connecting the Eq.(20), we get the entropy
\begin{eqnarray} \label{29}
S = \int {\frac{{d(1 - 8\pi {\eta ^2})M}}{{{T_H}}}}  = 4G{M^2}\pi {(1 - 8\pi {\eta ^2})^2}.
\end{eqnarray}
In order to study the phase transition of the Schwarzschild black hole with the global monopole, we get its free energy as follows \cite{31,32},
\begin{eqnarray} \label{30}
 F &&= {E_{Local}} - {T_{Local}}S \nonumber \\
  &&= \frac{{(1 - 8\pi {\eta ^2})\left[ {{M^2}( - 2 + 3{m^2}\beta ) + \beta {\hbar ^2}} \right]}}{{4M\sqrt {1 - \frac{{2M}}{r}} }} \nonumber \\
  &&- \frac{1}{2}\sqrt {\frac{{2( - 1 + 8\pi {\eta ^2})M + r}}{r}} \left[ {r(2 - 3{m^2}\beta ) + \frac{{\beta {\hbar ^2}}}{{( - 1 + 8\pi {\eta ^2})M}}} \right] \nonumber \\
  &&+ \frac{1}{2}\left[ {r(2 - 3{m^2}\beta ) + \frac{{\sqrt 2 \sqrt {\beta {\hbar ^2}} }}{{ - 1 + 8\pi {\eta ^2}}}} \right]\sqrt {\frac{{r + \sqrt 2 ( - 1 + 8\pi {\eta ^2})\sqrt {\beta {\hbar ^2}} }}{r}}  \nonumber \\
  &&+ \frac{{\beta {\hbar ^2}ArcTanh[\sqrt {\frac{{2( - 1 + 8\pi {\eta ^2})M + r}}{r}} ]}}{r} \nonumber \\
  &&- \frac{{\beta {\hbar ^2}ArcTanh[\sqrt {\frac{{r + \sqrt 2 ( - 1 + 8\pi {\eta ^2})\sqrt {\beta {\hbar ^2}} }}{r}} ]}}{r}.
\end{eqnarray}

\begin{figure}[h]
\centering
\includegraphics[width=0.6\textwidth]{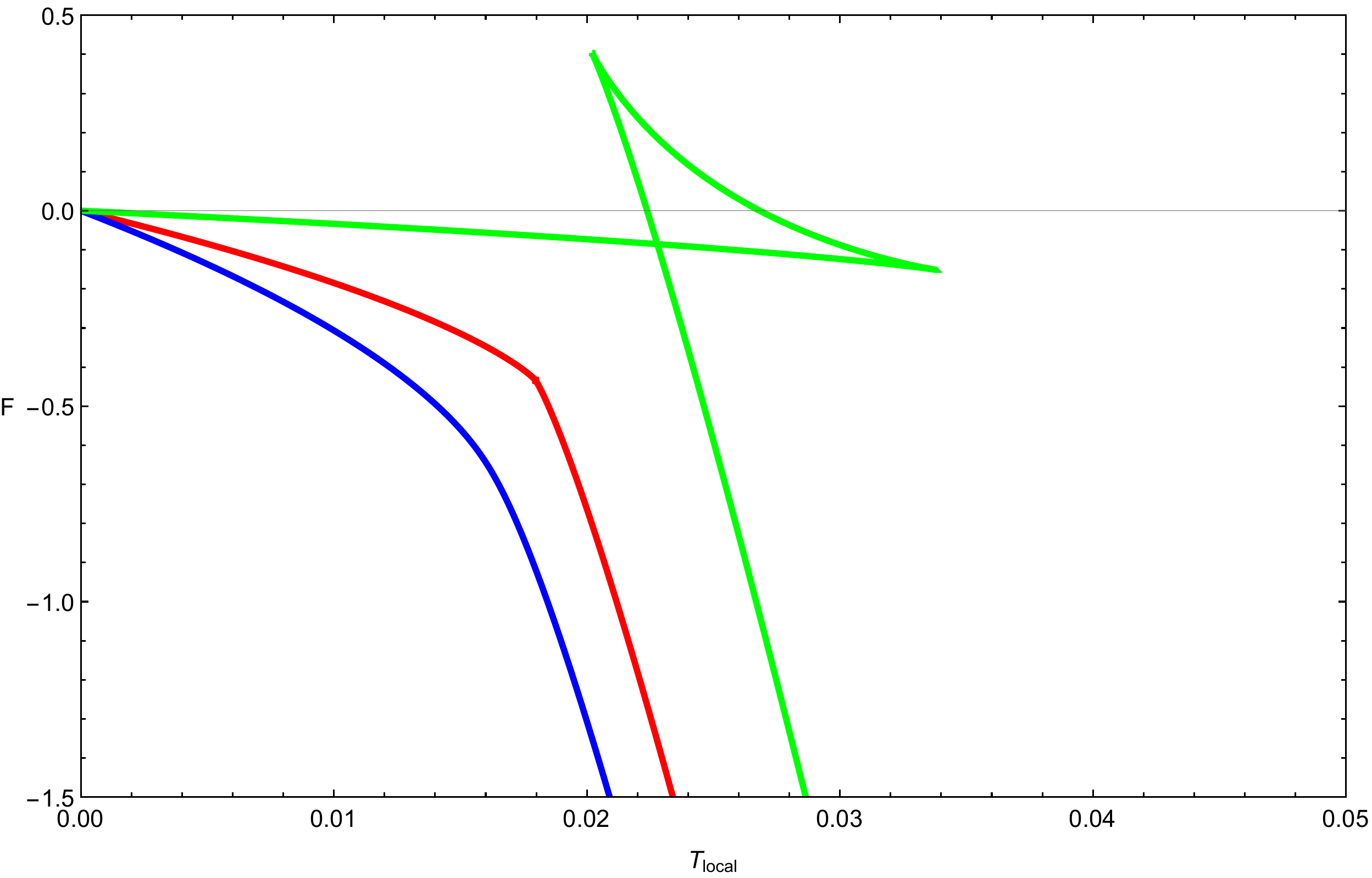}
\caption{\small Plot the relationship between local temperature  ${T_{Local}}$ and free energy $F$. The curve is the image from top to bottom for $\beta  = 4 > {\beta _C}$, $\beta  = {\beta _C}$, $\beta  = 0.5 < {\beta _C}$. The parameters are taken as $r = 10$, $\hbar  = 1$, $X = {10^{ - 5}}$.}
\end{figure}

From Figure 5, we show free energy $F$ curves with local temperature ${T_{Local}}$ under different generalized uncertainty parament $\beta $. It can be clearly seen that a dovetail structure appears, when the GUP parameter $\beta $  at this time is smaller than the critical value ${\beta _{Cr}}$, which means two phases coexist. That is the first order phase transition occurs. When the GUP parameter $\beta $ is equal to the critical value ${\beta _{Cr}}$, an inflection point occurs. The inflection point is the second-order phase transition point. When $\beta $ is more than ${\beta _{Cr}}$, the inflection point disappears. The trend of the image corresponds to the trend shown in the ${\rm{T}} - {\rm{M}}$ image in Figure 1.
To further explore the effect of $\beta $ on phase transitions. A plot of local temperature ${T_{Local}}$ and free energy F with $\beta  = 0$ is shown in Figure 6.

\begin{figure}[h]
\centering
\includegraphics[width=0.6\textwidth]{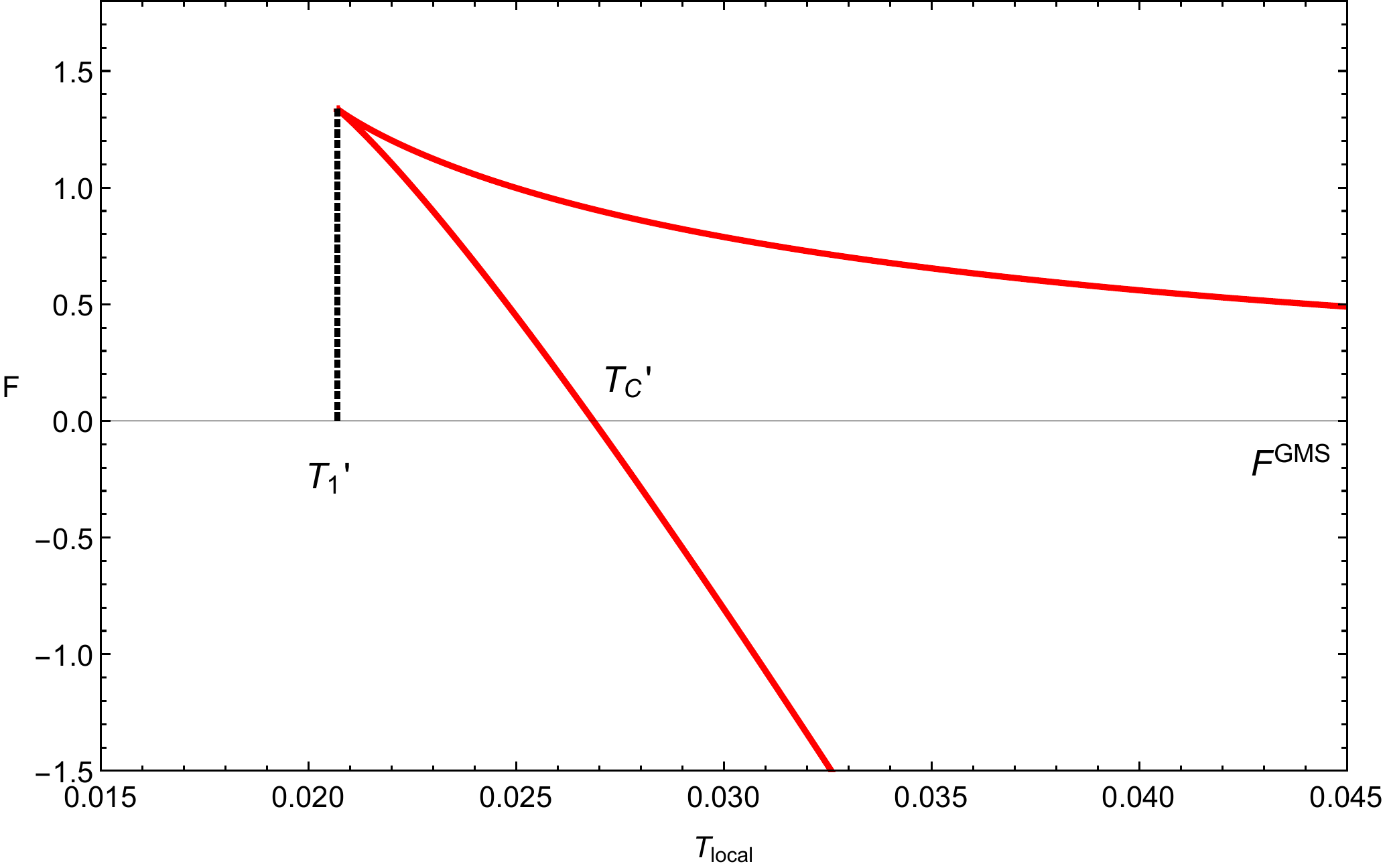}
\caption{\small Plot the local temperature ${T_{Local}}$ versus the free energy F for $\beta  = 0$. The curve corresponds to the phase transition. The parameters are taken as $r = 10$, $\hbar  = 1$, $X = {10^{ - 5}}$.}
\end{figure}

Figure 6 shows the relationship between free energy and local temperature in the original black hole. It can be found that the local temperature has a minimum at ${{\rm{T}}_1}'$ and there is no black hole solution below it. When $\beta  = 0$ and $M = 0$, black hole returns to monopole space-time. Therefore, ${F^{GMS}}$ represents the free energy of global monopole space time(GMS). For ${\rm{T }} < {{\rm{T}}_C}'$, the free energy of unstable black holes and stable large black holes is higher than ${F^{GMS}}$, which means that GMS is more popular than unstable black holes and stable large black holes. However, when ${\rm{T  > }}{{\rm{T}}_C}'$, the free energy of the unstable black hole is higher than ${F^{GMS}}$, and the free energy of the stable large black hole is lower than ${F^{GMS}}$. This indicates that the stable large black hole is more popular than GMS. So at ${\rm{T  > }}{{\rm{T}}_C}'$,  we can find that the phase transition tunneling radiation collapses to a stable large black hole, and the unstable small black hole eventually evolves into a large black hole.

\begin{figure}[h]
\centering
\includegraphics[width=0.6\textwidth]{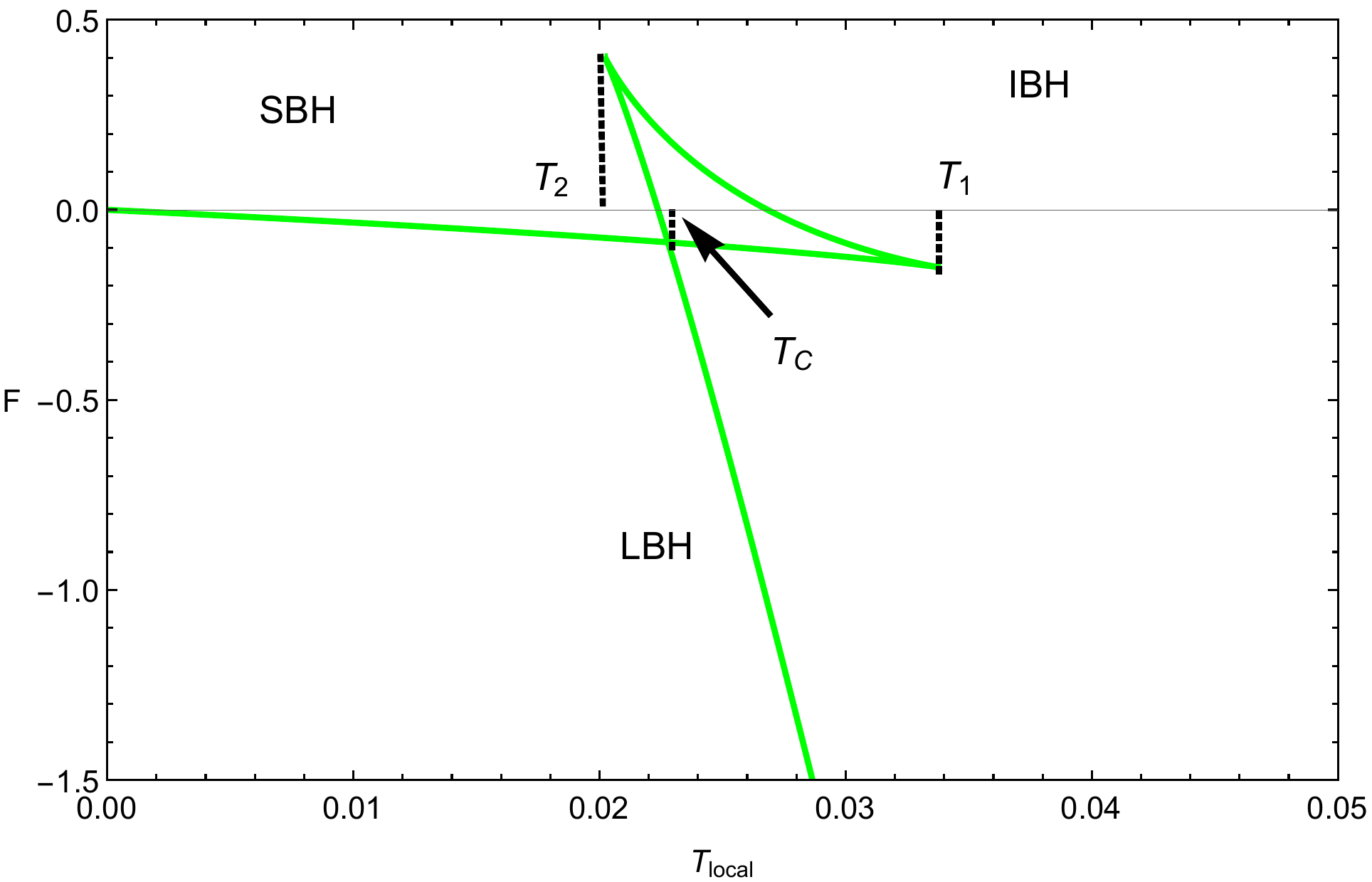}
\caption{\small Plot the relationship between local temperature ${T_{Local}}$ and free energy $F$ for $\beta  = 0$. The curve is the image for $\beta  = 0.5 < {\beta _C}$. The parameters are taken as $r = 10$, $\hbar  = 1$, $X = {10^{ - 5}}$.}
\end{figure}

As can be seen from Figure 7, F is free energy of curve space (CS). The SBH transforms to IBH at ${T_1}$, and the IBH transforms to LBH at ${T_2}$. The dovetail structure corresponds to a first-order phase transition. When ${T_0} < T < {T_C}$, ${F^{SBH}} < {F^{LBH}} < {F^{IBH}}$, which means that SBH is more possible than LBH and IBH, therefore LBH and IBH will decay into SBH; when ${T_c} < T < {T_1}$, ${F^{LBH}} < {F^{SBH}} < {F^{IBH}}$, which means that LBH is more possible than SBH and IBH, hence SBH and IBH will decay into LBH.

\section{Conclusion}

Based on GUP, we studies the phase transition and stability of Schwarzschild black holes with global monopoles. We calculated corrected thermodynamic quantities such as tunneling temperature, local temperature, heat capacity, and the entropy of the black hole. From the relationship between these thermodynamic quantities, we analyzed the phase transition and thermodynamic behavior of black holes. We find that GUP prevents black holes from completely evaporating and produces a stable residual mass, that is ${M_{res}} = \sqrt {\frac{{{\hbar ^2}\beta }}{2}} $. Secondly, when the GUP parameter $\beta $ and $M$ are close to zero, the heat capacity of the black hole disappears. When $\beta  < {\beta _C}$, the heat capacity of a Schwarzschild black hole with a monopole has two stable regions, and there is a first-order phase transition, that is, two phases coexist. Finally, according to the image of free energy vs. local temperature, there are two phase transitions. One is a dovetail structure, which means that there is a first-order phase transition in the Schwarzschild black hole with global monopoles at $\beta  < {\beta _C}$, and the other is a second-order phase transition at $\beta  = {\beta _C}$. In addition, we also found that with increasing temperature, stable SBH and unstable IBH will eventually decay into stable LBH. It is worth noting that the effects of GUP parameter and monopole parameter on the thermodynamic properties of Schwarzschild black holes are discussed with GUP parameter as variant. In the following work, we will discuss the thermodynamic behavior of a black hole with an integral magnetic monopole using the magnetic monopole parameters as variant.

\section*{Acknowledgements}

This work is supported in part by the National Natural Science Foundation of China (Grant No. 11703018), Natural Science
Foundation of Liaoning Province, China (Grant No. 20180550275) and Doctoral Scientific Research Foundation of Shenyang Normal University (Grant No. BS201843).

\end{document}